\documentclass[aps,prd,superscriptaddress,showpacs,preprintnumbers]{revtex4}

\newcommand{\be}{\begin{equation}}
\newcommand{\ee}{\end{equation}}
\newcommand{\bea}{\begin{eqnarray}}
\newcommand{\eea}{\end{eqnarray}}

\newcommand{\bfk}{\mbox{\boldmath $k$}}

\newcommand{\tSA}{S_{\ell A}}

\newcommand{\phip}{\phi^{\prime}}
\newcommand{\phig}{\phi_{\gamma}}
\newcommand{\phiSA}{\phi_{S_A}}
\newcommand{\thetap}{\theta^{\prime}}
\newcommand{\bfq}{\mbox{\boldmath $q$}}
\newcommand{\Xg}{\mbox{\boldmath $X$}_{\gamma}}
\newcommand{\Yg}{\mbox{\boldmath $Y$}_{\gamma}}
\newcommand{\Zg}{\mbox{\boldmath $Z$}_{\gamma}}
\newcommand{\Xcm}{\mbox{\boldmath $X$}_{cm}}
\newcommand{\Ycm}{\mbox{\boldmath $Y$}_{cm}}
\newcommand{\Zcm}{\mbox{\boldmath $Z$}_{cm}}
\newcommand{\kax}{k_{\perp a 1}}
\newcommand{\kay}{k_{\perp a 2}}
\newcommand{\gkk}{g(k_{\perp a}^2, \bfk_{\perp a}\cdot \bfq_{T})}
\newcommand{\bfS}{\mbox{\boldmath $S$}}
\newcommand{\bfzero}{\mbox{\boldmath $0$}}
\newcommand{\kpss}{\langle k_{\perp a}^2\rangle}
\def\S{_{_S}}
\begin{document}

%opening

%\maketitle
\title{Polarized and unpolarized Drell-Yan angular distribution \\ in the helicity formalism}
 \author{M.~Boglione}
 \affiliation{Dipartimento di Fisica Teorica, Universit\`a di Torino,
              Via P.~Giuria 1, I-10125 Torino, Italy\\
              INFN, Sezione di Torino, Via P.~Giuria 1, I-10125 Torino, Italy}
% %
 \author{S.~Melis}
 \affiliation{European Centre for Theoretical Studies in Nuclear Physics and Related Areas (ECT*) \\
              Villa Tambosi, Strada delle Tabarelle 286, I-38123 Villazzano, Trento, Italy}
\date{\today}
\begin{abstract}
We present a decomposition of the hadronic tensor for a general polarized Drell-Yan process
$AB\to \ell^{+}\ell^{-} X$ in terms of helicity structure functions using the helicity axes of the dilepton rest frame as a basis.
Next, we consider a QCD parton model and in the framework of a generalized QCD factorization scheme, which applies when the dilepton invariant mass $M$ is much larger than the transverse component of the photon momentum, $q_T$, in the hadronic center of mass frame. In this approximation we compute the angular distribution of the unpolarized Drell Yan cross section and of the Sivers effect, in single polarized Drell-Yan scattering, by taking into account the transverse motion of partons inside the initial hadrons, $k_\perp$. 
Interesting and simple results are found in the kinematical region $k_\perp \simeq q_T \ll M$, to first  order in a  $q_T / M$ expansion.
Finally, explicit analytical expressions, convenient for phenomenological studies, are obtained assuming a factorized Gaussian dependence on intrinsic momenta for the unpolarized and Sivers distribution functions, in analogy to those obtained for semi-inclusive deep inelastic scattering.
\end{abstract}

\maketitle

\section*{INTRODUCTION}

Lepton pair production in hadronic collisions, $AB\to \ell^{+}\ell^{-}X$, is one of the most useful tools to study  hadronic interactions and is presently the focus of considerable interest from both experimental and theoretical points of view.
Drell-Yan (DY) scattering processes in facts, due to their purely leptonic final state,
can be considered among the cleanest processes in order to study
the hadron structure,
including the puzzling question of the nucleon spin structure.
Moreover, at Born level, they are purely electromagnetic processes,
involving one elementary scattering channel ($q\bar{q}$ annihilation): therefore they play,
in hadronic interactions, the same, crucial role of DIS processes in electromagnetic interactions.

Here we present a decomposition of the hadronic tensor
for a general polarized DY process in terms of ``helicity'' structure functions.
Similar decompositions were presented in Ref.~\cite{Lam:1978pu} for the unpolarized case and in Ref.~\cite{Pire:1983tv} for the single polarized case. In Ref.~\cite{Arnold:2008kf}
the full expression for the general polarized DY cross section was obtained by using ``invariant'' structure functions, i.e.~using the spin vectors, the hadronic momenta and the dilepton momentum as a basis. This method is rather technical and lengthy.
Instead, we will show that the use of ``helicity'' structure functions, inspired by the approach presented in Ref.~\cite{Lam:1978pu}, is considerably simpler, elegant and easy to check.
Moreover the helicity structure functions can be directly
related to hadronic degrees of freedom,
i.e. they can be given a partonic interpretation.
Invariant structure functions and helicity structure functions
are obviously related to each other, as explained in details in Ref.~\cite{Lam:1978pu}.
We present some relevant results obtained in a non-collinear  extension of the QCD parton model framework, where we take into account the transverse motion of the constituents: the kinematical configuration of the process becomes much richer, allowing for 
%Considering the partonic transverse motion one has 
more degrees of
freedom and new possible correlations between hadron and parton spins and momenta.
Thus, beside the three usual collinear distribution functions
(unpolarized, helicity and transversity distribution functions),
five more Transverse Momentum Dependent distribution functions (TMDs) are necessary
to describe the nucleon structure.
Among them we find the Sivers function~\cite{Sivers:1989cc,Sivers:1990fh}, which gives the probability to find an unpolarized parton inside
a transversally polarized nucleon, and
the so called Boer-Mulders function describing
a transversally polarized parton inside an unpolarized nucleon.
The TMDs usually exhibit highly non-trivial QCD properties
(for instance, some of them are naively T-odd).
Their theoretical definition and their QCD evolution are still largely
unknown.
In spite of that, there are many experimental evidences
on the existence of these phenomena, provided
by several experimental collaborations in many facilities
in the world: HERMES at DESY, COMPASS at CERN, RHIC at BNL, BELLE at KEK, and JLAB (for a theoretical and experimental comprehensive review, see Ref.~\cite{Barone:2010zz} and references therein).

The inadequacy of the simple, collinear QCD  parton model emerging from the latest experimental data involving polarized hadronic reactions
motivates our efforts towards the study of a more refined picture, in which the ``transverse'', three-dimensional structure of hadrons relevantly comes into play, offering a fundamental ground for the investigation of the detailed form of intrinsic transverse momentum dependent (TMD) parton distribution functions (PDFs), which embed all the complexity of the above-mentioned 3-D structure of hadrons.

While in inclusive DIS the parton intrinsic transverse momentum of partons is invisible in any observable, as it is totally reabsorbed in the integration, this does not happen in processes like Drell-Yan or Semi-Inclusive DIS (SIDIS), where it becomes manifest in certain observables, like the angular distributions of the lepton pairs (i.e. the cross sections differential with respect to $q_T$ in DY) or the angular distributions of the final detected hadron (i.e.  the cross sections differential with respect to  $P_T$, the hadronic transverse momentum, in SIDIS).

In Sect.~\ref{UNP} we start by giving  a brief review of the general decomposition of the unpolarized DY cross section in terms of helicity structure functions~\cite{Lam:1978pu}, which we prefer to the invariant structure functions as they naturally show the hadronic structure properties in terms of their elementary constituents. While performing the few necessary steps to obtain the hadronic tensor decomposition, we set up a convenient methodology which will naturally lead our way through the much more complex polarized case.

In Sect.~\ref{S-POL} we define a covariant hadronic spin vector, and decompose the single polarized term of the hadronic tensor. We conclude with some interesting comments about the angular dependence of the single polarized DY cross section in two of the most used reference frames, the Collins-Soper (CS) dilepton rest frame~\cite{Collins:1977iv} and the hadronic center of mass (c.m.) frame.
Similarly, in Sect.~\ref{D-POL} we work out the decomposition of the doubly polarized term of the hadronic tensor and obtain the corresponding angular distributions.

In Sect.~\ref{P-M} we work in the QCD parton model approximation, where factorization allows us to write the DY hadronic tensor as the incoherent sum of the  partonic tensors of the corresponding parton annihilation processes. For the time being, we consider only the contributions to the DY cross section generated by {\it unpolarized} $q\bar q \to \ell ^+ \ell^-$ elementary cross section, i.e. the terms proportional to the convolution of two unpolarized TMD PDFs and to the Sivers effect. The remaining polarized terms will be considered elsewhere.
The full angular distribution of such terms is given in the CS frame, to first order in a $q_T/M$ expansion (where $q_T$ is the photon transverse momentum as measured in the hadronic c.m. frame).
Finally, the phenomenological implications of our results are discussed by considering a simple Gaussian model for the $k_\perp$ distribution of the TMDs.

Our conclusions are summarized in Sect.~\ref{CONCL}.

\section{Brief review of the unpolarized case}\label{UNP}

The general expression for the angular distribution of the polarized and unpolarized Drell-Yan process can be predicted by means of a kinematical analysis.
The cross section for the process $AB\to \ell^+ \ell^-X$ can be expressed in terms of the hadronic and leptonic tensors, $W^{\mu \nu}$ and $L^{\mu \nu}$ respectively,~\cite{Lam:1978pu,Boer:2006eq} as:
\begin{equation}
\frac{d\sigma}{d^4 q d\Omega}=\frac{\alpha^2}{2(2\pi^4)s M^4}L_{\mu\nu}W^{\mu\nu} \label{master_sez}
\end{equation}
where $q=(q_0,\bfq _T,q_L )$ is the dilepton momentum in the hadronic $c.m.$ frame, while $q^2=M^2$ is the dilepton mass squared.  The unpolarized, symmetric leptonic tensor is well known and can easily be written in term of the lepton momenta, $l_3$ and $l_4$:
\begin{equation}
 L^{\mu\nu}=2 l_3^{\mu}l_4^{\nu}+2 l_4^{\mu}l_3^{\nu}-M^2g^{\mu\nu}\label{leptonic_tensor_master}
\end{equation}
The hadronic tensor is a non-perturbative object describing the dynamic of the interaction between the two colliding hadrons. However, it is not  totally unknown as it must fulfill fundamental constraints like:
\begin{eqnarray}
q_{\mu}W^{\mu\nu}=q_{\nu}W^{\mu\nu}=0&\qquad&\textrm{current conservation;}\label{currentcons}\\
W^{\mu\nu}=W^{\nu\mu*}&\qquad&\textrm{hermiticity;}\label{hermit}\\
\mathcal{P}W^{\mu\nu}\mathcal{P}^{-1}=W^{\mu\nu}&\qquad&\textrm{parity invariance.}\label{parity}
\end{eqnarray}
where
$\mathcal{P}$ is the parity operator. Moreover, the hadronic tensor must be a function of the physical invariants of the process. Consequently, it is possible to decompose the hadronic tensor by means of structure functions, in complete analogy to what is usually done in deep inelastic processes. In particular here we want decompose the hadronic tensor by means of the so-called helicity structure functions, i.e. projecting the hadronic tensor on the components of the
virtual photon polarization vector~\cite{Lam:1978pu}.

For the unpolarized Drell-Yan process this leads to the well known decomposition in terms of the unpolarized helicity structure functions $W_T$, $W_L$, $W_{\Delta}$, $W_{\Delta\Delta}$~\cite{Lam:1978pu}:
\begin{eqnarray}
W^{\mu\nu}&=&\left(-g^{\mu\nu}+\frac{q^{\mu}q^{\nu}}{M^2}\right)(W_T+W_{\Delta\Delta})-2x^{\mu}x^{\nu} W_{\Delta\Delta}\nonumber\\
&+&z^{\mu}z^{\nu}(W_L-W_T-W_{\Delta\Delta})-(x^{\mu}z^{\nu}+z^{\mu}x^{\nu})W_{\Delta}\label{LamTungAngT}
\end{eqnarray}
where $W_T$ and $W_L$ are the structure functions for transversely and longitudinally polarized virtual photon, while $W_{\Delta}$ and $W_{\Delta\Delta}$ correspond to single and double spin flip respectively; $x^{\mu}$, $z^{\mu}$ (and $y^{\mu}$) are the dilepton rest frame axes. They are four-vectors defined in such a way that they are orthogonal to each other and to the four-vector $q^{\mu}$, and transform in unitary three-vectors in the dilepton rest frame. For convenience the axes are normalized to $-1$. An invariant expression of $x^{\mu}$, $y^{\mu}$, $z^{\mu}$ can be found in Ref.~\cite{Boer:2006eq}
% and appendix \ref{app-cs}
.
Notice that $W^{\mu}_{\mu}=-(2W_T+W_L)$, having followed the convention $g^{00}=1$, $g^{ii}=-1$ $(i=1,2,3)$ while  Ref.~\cite{Lam:1978pu} follows the opposite convention.
Contracting the hadronic tensor with the leptonic tensor, Eqs.~(\ref{leptonic_tensor_master}, \ref{leptonic_tensor}), one obtains the most general structure for the unpolarized Drell-Yan angular distribution:
\begin{eqnarray}
\label{LamTungAng}
\frac{d\sigma^{unp}}{d\Omega^{\prime}d^4 q}&=&\frac{\alpha^2}{2(2\pi)^4M^2 \,s}\Big[W_T(1+\cos^2\theta^{\prime})+W_L(1-\cos^2\theta^{\prime})\nonumber\\
&+&W_{\Delta} \sin 2\theta^{\prime} \cos\phi^{\prime}+W_{\Delta\Delta}\sin^2\theta^{\prime}\cos2\phi^{\prime}\Big]
\end{eqnarray}
where $\theta^{\prime}$ and $\phi^{\prime}$ are respectively the lepton polar and azimuthal angles in the dilepton rest frame defined by the chosen axes $x^{\mu}$, $y^{\mu}$, $z^{\mu}$.
A similar expression for the single polarized case can be found in Ref.~\cite{Pire:1983tv} while Ref.~\cite{Arnold:2008kf} shows the most general case where both hadrons can be polarized. Here we present a different, elegant derivation using the properties of the hadronic tensor,~Eqs.~(\ref{currentcons},\ref{hermit},\ref{parity}), and a decomposition of this tensor in terms of helicity structure functions projected onto the helicity axes of the dilepton rest frame, in complete analogy to the unpolarized case outlined above.
In order to explore the polarized case it is better to use the more flexible notation of Ref.~\cite{Arnold:2008kf}. For example, we rewrite Eq.~(\ref{LamTungAngT}) in this way:
\begin{eqnarray}
W^{\mu\nu}&=&\left(-g^{\mu\nu}+\frac{q^{\mu}q^{\nu}}{M^2}\right)\left(F_{UU}^1+F_{UU}^{\cos2\phip}\right)-2x^{\mu}x^{\nu} F_{UU}^{\cos2\phip}\nonumber\\
&&+z^{\mu}z^{\nu}\left(F_{UU}^2-F_{UU}^1-F_{UU}^{\cos2\phip}\right)-\left(x^{\mu}z^{\nu}+z^{\mu}x^{\nu}\right)F_{UU}^{\cos\phip}
\label{LamTungAM}
\end{eqnarray}
where $UU$ denotes that both hadrons are unpolarized and $F_{UU}^1\equiv W_T$, $F_{UU}^2\equiv W_L$, $F_{UU}^{\cos\phi^{\prime}}\equiv W_{\Delta}$ and $F_{UU}^{\cos2\phi^{\prime}}\equiv W_{\Delta\Delta}$.
This notation has the advantage of reminding the azimuthal dependence of each helicity structure function.

\section{Single polarized Drell-Yan}\label{S-POL}

To deal with the polarized Drell Yan processes, we need to properly take into account the polarization degrees of freedom. For this reason, it is convenient to define the following four-vectors:
\begin{eqnarray}
S^{\mu}_h&=&(S_{0h},\bfS_{Th},S_{Lh})\\
S^{\mu}_{Th}&=&(0,\bfS_{Th},0)\\
S^{\mu}_{Lh}&=&(S_{0h},\bfzero,S_{Lh})\\
S^{\mu}_{\ell h}&=&(z\cdot S_{h}) z^{\mu} \\
S_{\perp h}^{\mu}&=&S^{\mu}_h-\Big(\frac{q^\mu q^\nu}{M^2}+z^{\mu} z^{\nu}\Big)S_{h\nu}\,,
\end{eqnarray}
where $S^{\mu}_h=(S_{0h},\bfS_{Th},S_{Lh})$ is the covariant spin vector of hadron $h=A,B$ in the hadronic c.m. frame. It can be easily seen that $S^{\mu}_{\ell h} q_{\mu}=0$, $S_{\perp h}^{\mu}q_{\mu}=0$. This property is very important because any decomposition of the hadronic tensor by means of the helicity axes $x^{\mu}$, $y^{\mu}$, $z^{\mu}$ and the $S^{\mu}_{\ell h}$ and $S_{\perp h}^{\mu}$ vectors automatically fulfills the requirement of Eq.~(\ref{currentcons}). Notice also that
$S_{\perp h}^{\mu}$ and $S^{\mu}_{\ell h}$ are purely transverse $(0,\bfS^\prime_{\perp h},0)$ or longitudinal $(0,0,0,S^\prime_{\ell h})$ (three-vectors) in the dileptonic rest frame.

In general, the hadronic polarized tensor can be written as a sum of four terms:
%
%\begin{equation}
%W^{\mu\nu}(S_A)= W^{\mu\nu}_{unp}+W^{\mu\nu}_{S_A}+W^{\mu\nu}_{S_B}+W^{\mu\nu}_{S_{A} S_{B}}\,,
%\end{equation}
%
\begin{equation}
W^{\mu\nu}(S_A,S_B)= W^{\mu\nu}_{unp}+W^{\mu\nu}_{S_A}+W^{\mu\nu}_{S_B}+W^{\mu\nu}_{S_{A} S_{B}}\,,\label{eq:WSASB}
\end{equation}
where $W^{\mu\nu}_{unp}$ is the unpolarized hadronic tensor of Eqs.~(\ref{LamTungAngT}) or (\ref{LamTungAM}); $W^{\mu\nu}_{S_A}$ and $W^{\mu\nu}_{S_B}$ are the single polarized terms of the hadronic tensor while $W^{\mu\nu}_{S_A S_B}$ is the double polarized term.

%\subsection{Single polarized Drell-Yan}

Applying current conservation, hermiticity and parity conservation, Eqs.~(\ref{currentcons}-\ref{parity}), the single polarized hadronic tensor can be decomposed in this way:
\begin{eqnarray}
W^{\mu\nu}_{S_A}&=&-\left(S_{\perp A}\cdot y\right)\left(F_{\perp U}^1+F_{\perp U}^{\cos2\phip}\right)
\left(-g^{\mu\nu}+\frac{q^{\mu}q^{\nu}}{M^2}\right)
\nonumber\\
&+&2\left(S_{\perp A}\cdot y\right)\left(F_{\perp U}^{\cos2\phip}+F_{\perp U}^{\sin2\phip}\right)x^{\mu}x^{\nu}
\nonumber\\
&+&\left(S_{\perp A}\cdot y\right)\left(F_{\perp U}^1-F_{\perp U}^2+F_{\perp U}^{\cos2\phip}\right)z^{\mu}z^{\nu}\nonumber\\
&+&\left(S_{\perp A}\cdot y\right)F_{\perp U}^{\cos\phip}(x^{\mu}z^{\nu}+z^{\mu}x^{\nu})\nonumber\\
&-&F_{\perp U}^{\sin2\phip}(S_{\perp A}^{\mu}y^{\nu}+y^{\mu}S_{\perp A}^{\nu})
-F_{\ell U}^{\sin\phip}(\tSA^{\mu}y^{\nu}+y^{\mu}\tSA^{\nu})\nonumber\\
&+&\left(\tSA\cdot z\right)F_{\ell U}^{\sin2\phip}(x^{\mu}y^{\nu}+y^{\mu}x^{\nu})\nonumber\\
&+&\left(S_{\perp A}\cdot x\right) F_{\perp U}^{\sin\phip}(z^{\mu}y^{\nu}+y^{\mu}z^{\nu})\label{eq:singlepW}\,.
\end{eqnarray}
In the first four lines we recognize the the same patterns found for the unpolarized cross section of Eq.~(\ref{LamTungAM}), while in the last three lines we find genuinely new structures, proportional to the spin vector components. For the $F$ structure functions we have adopted a notation similar to that used in Eq.~(\ref{LamTungAM}), although their explicit connection to the stated angular dependence will only become known once the contraction with the leptonic tensor has been performed.

The general, single polarized DY cross section can conveniently be separated in two parts
\begin{equation}
\frac{d\sigma(S_A)}{d\Omega^{\prime}d^4 q}=\frac{d\sigma^{unp}}{d\Omega^{\prime}d^4 q}+\frac{d\sigma^{S_A}}{d\Omega^{\prime}d^4 q}\,,
\end{equation}
where $\frac{d\sigma^{unp}}{d\Omega^{\prime}d^4 q}$ is the unpolarized cross section of Eq.~(\ref{LamTungAngT}).
Contracting the single polarized hadronic tensor of Eq.~(\ref{eq:singlepW}) with the leptonic tensor of Eq.~(\ref{leptonic_tensor}), according to Eq.~(\ref{master_sez}), we obtain the polarized term of the above cross section
\begin{eqnarray}
\frac{d\sigma^{S_A}}{d\Omega^{\prime}d^4 q}&=&
\frac{\alpha^2}{2(2\pi)^4M^2\,s}\nonumber\\
&&\times\Bigg\{S_{\perp A}^{\prime}\sin\phi_{S_{A}}^{\prime}\Big[F_{\perp U}^1(1+\cos^2\theta^{\prime})+ F_{\perp U}^2(1-\cos^2\theta^{\prime})\nonumber\\
&&\qquad\qquad\qquad\;\;\;\;+ F_{\perp U}^{\cos\phip}\sin 2\theta^{\prime} \cos\phi^{\prime}+F_{\perp U}^{\cos2\phip}\sin^2\theta^{\prime}\cos2\phi^{\prime}\Big]\nonumber\\
&&\quad+S_{\perp A}^{\prime}\cos\phi_{S_{A}}^{\prime}\Big[F_{\perp U}^{\sin\phip}\sin 2\theta^{\prime}\sin\phi^{\prime}+F_{\perp U}^{\sin2\phip}\sin^2\theta^{\prime}\sin2\phi^{\prime}\Big]\nonumber\\
&&\quad+S_{\ell A}^{\prime} \Big[F_{\ell U}^{\sin\phip}\sin 2\theta^{\prime}\sin\phi^{\prime}+F_{\ell U}^{\sin2\phip}\sin^2\theta^{\prime}\sin2\phi^{\prime}\Big]\Bigg\}\,.
\label{eq:singlepxs}
\end{eqnarray}
Expressions similar to Eqs.~(\ref{eq:singlepW}) and (\ref{eq:singlepxs}) can easily be written for polarized hadron $B$.
Here the primes denote that the angles are in the leptonic c.m. frame, analogously the polarization directions $\ell$ and $\perp$ in the $F$ structure functions are referred to the dilepton rest frame, and
\begin{eqnarray}
\left(S_{\perp A}^{\prime}\cos\phi_{S_{A}}^{\prime}\right)&\equiv&-\left(S_{\perp A}\cdot x\right) \\
\left(S_{\perp A}^{\prime}\sin\phi_{S_{A}}^{\prime}\right)&\equiv&-\left(S_{\perp A}\cdot y\right)\,.
\end{eqnarray}
Notice that there is a straightforward correspondence between the polarization directions $\ell$ and $\perp$ in the leptonic rest frame and  the longitudinal and transverse polarization directions, $L$ and $T$, in the hadronic c.m. frame. In particular
\begin{equation}
\left(S_{\perp A}^{\prime}\sin\phi_{S_{A}}^{\prime}\right)=-S_{\perp A}^{\mu}y_{\mu}=-S^{\mu}_{TA}y_{\mu}=S_{TA}\sin(\phi_{S_A}-\phi_{\gamma})
\,,
\label{eq:stymu}
\end{equation}
whereas
\begin{eqnarray}
S_{\perp A}^{\mu} x_{\mu} = S^{\mu}_{A}x_{\mu} &=& S_{TA}^{\mu}x_{\mu}+S_{LA}^{\mu}x_{\mu}\nonumber \\
S_{\ell A}^{\mu} z_{\mu}  = -S^{\mu}_{A}z_{\mu}&=&-S_{TA}^{\mu}z_{\mu}-S_{LA}^{\mu}z_{\mu} \,, \label{eq:SxSz}
\end{eqnarray}
and consequently
\begin{eqnarray}
\left(S_{\perp A}^{\prime}\cos\phi_{S_{A}}^{\prime}\right)&\ne& S_T\cos(\phi_{S_A}-\phi_{\gamma})\label{eq:stxmu}\\
S^\prime_{\ell A} &\ne& S_{LA}\,
\end{eqnarray}
where $\phi_{\gamma}$ and $\phi_{S_A}$ are the dilepton and the proton $A$ polarization azimuthal angles in the hadronic c.m. frame.
Notice that Eq.~(\ref{eq:stymu}) holds because by definition the $y^{\mu}$ axis is purely transverse in the hadronic c.m. frame, while $x^{\mu}$ and $z^{\mu}$ are not. Therefore a transverse spin vector in the dilepton rest frame can acquire a longitudinal component in the hadronic c.m. frame  (and, vice-versa, a longitudinal spin vector in the dilepton rest frame can acquire a transverse component in the hadronic c.m. frame).
This was also pointed out in Ref.~\cite{Arnold:2008kf}. Similar expressions can be found in a slightly different notation in Ref.~\cite{Pire:1983tv}.

For practical purposes, it is more useful to rewrite the polarized cross section of Eq.~(\ref{eq:singlepxs}) in terms of  polarizations which are purely transverse and longitudinal in the hadronic c.m. frame. This can be done inserting   Eqs.~(\ref{eq:stymu}-\ref{eq:SxSz}) into the general expression of the hadronic polarized tensor $W^{\mu\nu}_{S_A}$ of  Eq.~(\ref{eq:singlepW}) and rearranging the structure functions. Here, we show how this can be done adopting the Collins Soper (CS) frame~\cite{Collins:1977iv}, in which the kinematics is simple and transparent allowing to perform some useful approximations. Using the definitions of the CS axes $x_{\mu}$, $y_{\mu}$, $z_{\mu}$ as given in Appendix~\ref{app-cs} we obtain
\begin{eqnarray}
S_{TA}^{\mu} x_{\mu} &=& -\frac{\sqrt{M^2+q_T^2}}{M} S_{TA} \cos(\phi_{S_A}-\phi_{\gamma})\nonumber\\
S_{TA}^{\mu} y_{\mu} &=&-S_{TA} \sin(\phi_{S_A}-\phi_{\gamma})\nonumber\\
S_{TA}^{\mu} z_{\mu} &=& 0\nonumber\\
S_{LA}^{\mu} x_{\mu} &=& -\frac{q_T}{M}\frac{(q_0-q_L)}{\sqrt{M^2+q_T^2}} S_{LA} \nonumber\\
S_{LA}^{\mu} y_{\mu} &=&0\nonumber\\
S_{LA}^{\mu} z_{\mu} &=& -\frac{(q_0-q_L)}{\sqrt{M^2+q_T^2}} S_{LA}\,,
\end{eqnarray}
where $q_T$ and $q_L$ are, respectively, the transverse and longitudinal components of the virtual photon momentum in the hadronic c.m. frame.
With these in mind, we can write the single polarized Drell Yan cross section as
\begin{eqnarray}
\frac{d\sigma^{S_A}}{d\Omega^{\prime}d^4 q}&=&
\frac{\alpha^2}{2(2\pi)^4M^2\,s}\nonumber\\
&&\times\Bigg\{S_{TA}\sin\phi_{S_{A}}\Big[F_{TU}^1(1+\cos^2\theta^{\prime})+ F_{TU}^2(1-\cos^2\theta^{\prime})\Big]\nonumber\\
&&\;\;\;\;+ S_{TA} F_{TU}^{\sin(\phi_{S_A} - \phi_{\gamma} -\phip)}\sin 2\theta^{\prime} \sin(\phi_{S_A} - \phi_{\gamma} -\phip)\nonumber\\
&&\;\;\;\;+ S_{TA} F_{TU}^{\sin(\phi_{S_A} - \phi_{\gamma} +\phip)}\sin 2\theta^{\prime} \sin(\phi_{S_A} - \phi_{\gamma} +\phip) \nonumber\\
&&\;\;\;\;+ S_{TA} F_{TU}^{\sin(\phi_{S_A} - \phi_{\gamma} -2\phip)}\sin^2\theta^{\prime} \sin(\phi_{S_A} - \phi_{\gamma} -2\phip)\nonumber\\
&&\;\;\;\;+ S_{TA} F_{TU}^{\sin(\phi_{S_A} - \phi_{\gamma} +2\phip)}\sin^2\theta^{\prime} \sin(\phi_{S_A} - \phi_{\gamma} +2\phip) \nonumber\\
&&\quad+S_{LA} \Big[F_{LU}^{\sin\phip}\sin 2\theta^{\prime}\sin\phi^{\prime}
+F_{L U}^{\sin2\phip}\sin^2\theta^{\prime}\sin2\phi^{\prime}\Big]\Bigg\}\,,
\label{eq:singlepxs2}
\end{eqnarray}
having rearranged the $F$ structure functions in the following way
\begin{eqnarray}
F_{TU}^{\sin(\phi_{S_A} - \phi_{\gamma} \mp \phip)} &=&
F_{\perp U}^{\cos\phip} \pm \frac{\sqrt{M^2+q_T^2}}{M} F_{\perp U}^{\sin\phip}\nonumber\\
F_{TU}^{\sin(\phi_{S_A} - \phi_{\gamma} \mp 2\phip)} &=&
F_{\perp U}^{\cos2\phip} \pm \frac{\sqrt{M^2+q_T^2}}{M} F_{\perp U}^{\sin2\phip}\nonumber\\
F_{LU}^{\sin\phip} &=& \frac{(q_0-q_L)}{\sqrt{M^2+q_T^2}} \Big[\frac{q_T}{M} F_{\perp U}^{\sin\phip} - F_{\ell U}^{\sin\phip}\Big]\nonumber\\
F_{LU}^{\sin2\phip} &=& \frac{(q_0-q_L)}{\sqrt{M^2+q_T^2}} \Big[\frac{q_T}{M} F_{\perp U}^{\sin2\phip} - F_{\ell U}^{\sin2\phip}\Big]\,.
\end{eqnarray}
It is interesting to notice that, in the Collins-Soper frame, in the limit $q_T\ll M$, one has
$\left(S_{\perp A}^{\prime}\cos\phi_{S_{A}}^{\prime}\right)\simeq S_T\cos(\phi_{S_A}-\phi_{\gamma})$, i.e. $S_{\perp A} \simeq S_{TA}$,
and $S_{\ell A} \simeq S_{LA}$. Therefore, in this limit, Eq.~(\ref{eq:singlepxs}) holds with  $\perp\equiv T$
and $\ell \equiv L$, recovering the results of Ref.~\cite{Arnold:2008kf}, with $\phi_{\gamma}\equiv0$.

\section{Double polarized Drell-Yan} \label{D-POL}

For the doubly polarized Drell Yan process we proceed in a similar way. For convenience, we decompose the double polarized hadronic tensor $W^{\mu\nu}_{S_A S_B}$ in four parts, according to the longitudinal and/or transverse components of the spin vectors $S_A$ and $S_B$ which they involve:
\begin{equation}
 W^{\mu\nu}_{S_A S_B}=
W^{\mu\nu}_{S_{\ell A}^{\prime} S_{\ell B}^{\prime}} +
W^{\mu\nu}_{S_{\perp A}^{\prime} S_{\perp B}^{\prime}} +
W^{\mu\nu}_{S_{\ell A}^{\prime} S_{\perp B}^{\prime}} +
W^{\mu\nu}_{S_{\perp A}^{\prime} S_{\ell B}^{\prime}}\,.
\label{doubleW}
\end{equation}
By requiring current conservation, hermiticity and parity conservation, Eqs.~(\ref{currentcons}-\ref{parity}), we obtain the decomposition of each term as a function of $F$ helicity structure functions:

\begin{eqnarray}
W^{\mu\nu}_{S_{\ell A}^{\prime} S_{\ell B}^{\prime}}&=&
%doppia longitudinale
\left(S_{\ell A}\cdot z\right)\left(S_{\ell B}\cdot z\right)\left(F_{\ell\ell}^1+F_{\ell\ell}^{\cos2\phip}\right)
\left(-g^{\mu\nu}+\frac{q^{\mu}q^{\nu}}{M^2}\right)
\nonumber\\
&-&2\left(S_{\ell A}\cdot z\right)\left(S_{\ell B}\cdot z\right)\left(F_{\ell\ell}^{\cos2\phip}+F_{\ell\ell}^{\sin2\phip}\right)x^{\mu}x^{\nu}
\nonumber\\
&+&\left(S_{\ell A}\cdot z\right)\left(S_{\ell B}\cdot z\right)\left(F_{\ell\ell}^2-F_{\ell\ell}^1-F_{\ell\ell}^{\cos2\phip}\right)z^{\mu}z^{\nu}\nonumber\\
&-&\left(S_{\ell A}\cdot z\right)\left(S_{\ell B}\cdot z\right) F_{\ell\ell}^{\cos\phip}(x^{\mu}z^{\nu}+z^{\mu}x^{\nu})\,,
\label{eq:WSALSBL}
\end{eqnarray}

\begin{eqnarray}
%doppia trasversa
%cos phisa+phisb
W^{\mu\nu}_{S_{\perp A}^{\prime} S_{\perp B}^{\prime}}&=&
\left[\left(S_{\perp A}\cdot x\right)\left(S_{\perp B}\cdot x\right)-\left(S_{\perp A}\cdot y\right)\left(S_{\perp B}\cdot y\right)\right]\left(F_{\perp \perp}^1+F_{\perp \perp}^{\cos2\phip}\right)
\left(-g^{\mu\nu}+\frac{q^{\mu}q^{\nu}}{M^2}\right)
\nonumber\\
&-&2\left[\left(S_{\perp A}\cdot x\right)\left(S_{\perp B}\cdot x\right)-\left(S_{\perp A}\cdot y\right)\left(S_{\perp B}\cdot y\right)\right]\left(F_{\perp \perp}^{\cos2\phip}+F_{\perp \perp}^{\sin2\phip}\right)x^{\mu}x^{\nu}
\nonumber\\
&+&\left[\left(S_{\perp A}\cdot x\right)\left(S_{\perp B}\cdot x\right)-\left(S_{\perp A}\cdot y\right)\left(S_{\perp B}\cdot y\right)\right]\left(F_{\perp \perp}^2-F_{\perp \perp}^1-F_{\perp \perp}^{\cos2\phip}\right)z^{\mu}z^{\nu}\nonumber\\
&-& \left[\left(S_{\perp A}\cdot x\right)\left(S_{\perp B}\cdot x\right)-\left(S_{\perp A}\cdot y\right)\left(S_{\perp B}\cdot y\right)\right] F_{\perp \perp}^{\cos\phip}(x^{\mu}z^{\nu}+z^{\mu}x^{\nu})\nonumber\\
%
%cos phisa-phisb
&+&\left[\left(S_{\perp A}\cdot x\right)\left(S_{\perp B}\cdot x\right)+\left(S_{\perp A}\cdot y\right)\left(S_{\perp B}\cdot y\right)\right]\left(\bar{F}_{\perp\perp}^1+\bar{F}_{}^{\cos2\phip}\right)
\left(-g^{\mu\nu}+\frac{q^{\mu}q^{\nu}}{M^2}\right)
\nonumber\\
&-&2\left[\left(S_{\perp A}\cdot x\right)\left(S_{\perp B}\cdot x\right)+\left(S_{\perp A}\cdot y\right)\left(S_{\perp B}\cdot y\right)\right]\left(\bar{F}_{\perp\perp}^{\cos2\phip}+\bar{F}_{\perp\perp}^{\sin2\phip}\right)x^{\mu}x^{\nu}
\nonumber\\
&+&\left[\left(S_{\perp A}\cdot x\right)\left(S_{\perp B}\cdot x\right)+\left(S_{\perp A}\cdot y\right)\left(S_{\perp B}\cdot y\right)\right]\left(\bar{F}_{\perp\perp}^2-\bar{F}_{\perp\perp}^1-\bar{F}_{\perp\perp}^{\cos2\phip}\right)z^{\mu}z^{\nu}\nonumber\\
&-& \left[\left(S_{\perp A}\cdot x\right)\left(S_{\perp B}\cdot x\right)+\left(S_{\perp A}\cdot y\right)\left(S_{\perp B}\cdot y\right)\right] \bar{F}_{\perp\perp}^{\cos\phip}(x^{\mu}z^{\nu}+z^{\mu}x^{\nu})\nonumber\\
%
%sin phisa+phisb
&-&\left[\left(S_{\perp A}\cdot y\right)\left(S_{\perp B}\cdot x\right)+\left(S_{\perp A}\cdot x\right)\left(S_{\perp B}\cdot y\right)\right]F_{\perp \perp}^{\sin\phip}(z^{\mu}y^{\nu}+y^{\mu}z^{\nu})\nonumber\\
&-&\left[\left(S_{\perp A}\cdot y\right)\left(S_{\perp B}\cdot x\right)+\left(S_{\perp A}\cdot x\right)\left(S_{\perp B}\cdot y\right)\right]F_{\perp \perp}^{\sin2\phip}(x^{\mu}y^{\nu}+y^{\mu}x^{\nu})\nonumber\\
%sin phisa-phisb
&-&\left[\left(S_{\perp A}\cdot y\right)\left(S_{\perp B}\cdot x\right)-\left(S_{\perp A}\cdot x\right)\left(S_{\perp B}\cdot y\right)\right]\bar{F}_{\perp\perp}^{\sin\phip}(z^{\mu}y^{\nu}+y^{\mu}z^{\nu})\nonumber\\
&-&\left[\left(S_{\perp A}\cdot y\right)\left(S_{\perp B}\cdot x\right)-\left(S_{\perp A}\cdot x\right)\left(S_{\perp B}\cdot y\right)\right]\bar{F}_{\perp\perp}^{\sin2\phip}(x^{\mu}y^{\nu}+y^{\mu}x^{\nu})\label{eq:WSAperpSBPerp}
%-F_{\ell U}^{\sin\phip}(\tSA^{\mu}y^{\nu}+y^{\mu}\tSA^{\nu})\nonumber\\
%&+&\left(\tSA\cdot Z\right)F_{\ell U}^{\sin2\phip}(x^{\mu}y^{\nu}+y^{\mu}x^{\nu})\nonumber\\
%&+&\left(S_{\perp A}\cdot x\right) F_{\perp U}^{\sin\phip}(z^{\mu}y^{\nu}+y^{\mu}z^{\nu})%\label{eq:singlepW}\,,
\end{eqnarray}

\begin{eqnarray}
 W^{\mu\nu}_{S_{\ell A}^{\prime} S_{\perp B}^{\prime}}&=&\left(S_{\ell A}\cdot z\right)\left(S_{\perp B}\cdot x\right)\left(F_{\ell \perp}^1+F_{\ell \perp}^{\cos2\phip}\right)
\left(-g^{\mu\nu}+\frac{q^{\mu}q^{\nu}}{M^2}\right)
\nonumber\\
&-&2\left(S_{\ell A}\cdot z\right)\left(S_{\perp B}\cdot x\right)\left(F_{\ell \perp}^{\cos2\phip}+F_{\ell \perp}^{\sin2\phip}\right)x^{\mu}x^{\nu}
\nonumber\\
&+&\left(S_{\ell A}\cdot z\right)\left(S_{\perp B}\cdot x\right)\left(F_{\ell \perp}^2-F_{\ell \perp}^1-F_{\ell \perp}^{\cos2\phip}\right)z^{\mu}z^{\nu}\nonumber\\
&-&\left(S_{\ell A}\cdot z\right)\left(S_{\perp B}\cdot x\right) F_{\ell \perp}^{\cos\phip}(x^{\mu}z^{\nu}+z^{\mu}x^{\nu})\nonumber\\
&+&\left(S_{\perp B}\cdot y\right)F_{\ell \perp}^{\sin\phi}\left(S^{\mu}_{\ell A} y^{\nu}+y^{\mu}S^{\nu}_{\ell A}\right)\nonumber\\
&+&\left(S_{\ell A}\cdot z\right)F_{\ell \perp}^{\sin2\phi}\left(S^{\mu}_{\perp B} x^{\nu}+x^{\mu}S^{\nu}_{\perp B}\right)\,.
\label{eq:WSALSBPerp}
\end{eqnarray}

An analogous relation holds for $W^{\mu\nu}_{S_{\perp A}^{\prime} S_{\ell B}^{\prime}}$. Notice that, although the complexity that a doubly polarized hadron-hadron scattering inevitably presents, at a careful look we can clearly recognize, in each hadronic tensor term, the structures already found in the unpolarized and single polarized cases, Eqs.~(\ref{LamTungAM}) and (\ref{eq:singlepW}).

The general, doubly polarized DY cross section can conveniently be separated in four parts
\begin{equation}
 \frac{d\sigma(S_A S_B)}{d\Omega^{\prime}d^4 q}=
 \frac{d\sigma^{unp}}{d\Omega^{\prime}d^4 q}
 +\frac{d\sigma^{S_A}}{d\Omega^{\prime}d^4 q}
 +\frac{d\sigma^{S_B}}{d\Omega^{\prime}d^4 q}
 +\frac{d\sigma^{S_A S_B}}{d\Omega^{\prime}d^4 q}\,
\end{equation}
where $\frac{d\sigma^{unp}}{d\Omega^{\prime}d^4 q}$ is the unpolarized cross section of Eq.~(\ref{LamTungAngT}), while $\frac{d\sigma^{S_A}}{d\Omega^{\prime}d^4 q}$ and $\frac{d\sigma^{S_B}}{d\Omega^{\prime}d^4 q}$ are the single polarized contributions of Eq.~(\ref{eq:singlepxs}).
Contracting Eqs.~(\ref{eq:WSALSBL}-\ref{eq:WSALSBPerp}) with the leptonic tensor of Eq.~(\ref{leptonic_tensor}) we can compute the doubly polarized term of the above cross section:
\begin{eqnarray}
 \frac{d\sigma^{S_A S_B}}{d\Omega^{\prime}d^4 q}&=&\frac{\alpha^2}{2(2\pi)^4M^2\,s}\nonumber\\
%LL
&&\times\Bigg\{S_{\ell A}^{\prime}S_{\ell B}^{\prime}\Big[F_{\ell\ell}^1(1+\cos^2\theta^{\prime})+ F_{\ell\ell}^2(1-\cos^2\theta^{\prime})
\nonumber\\
&&\qquad\qquad\qquad\;\;\;\;+ F_{\ell\ell}^{\cos\phip}\sin 2\theta^{\prime} \cos\phi^{\prime}+F_{\ell\ell}^{\cos2\phip}\sin^2\theta^{\prime}\cos2\phi^{\prime}\Big]\nonumber\\
%
%\perp\perp
%cos phisa+phisb
&&\quad\;+S_{\perp A}^{\prime}S_{\perp B}^{\prime}\cos(\phi_{S_A}^{\prime}+\phi_{S_B}^{\prime})
\Big[F_{\perp \perp}^1(1+\cos^2\theta^{\prime})+ F_{\perp \perp}^2(1-\cos^2\theta^{\prime})\nonumber\\
&&\qquad\qquad\qquad\;\;\;\;+ F_{\perp \perp}^{\cos\phip}\sin 2\theta^{\prime} \cos\phi^{\prime}+F_{\perp \perp}^{\cos2\phip}\sin^2\theta^{\prime}\cos2\phi^{\prime}\Big]\nonumber\\
%
%
%cos phisa-phisb
&&\quad\;+S_{\perp A}^{\prime}S_{\perp B}^{\prime}\cos(\phi_{S_A}^{\prime}-\phi_{S_B}^{\prime})
\Big[\bar{F}_{\perp\perp}^1(1+\cos^2\theta^{\prime})+ \bar{F}_{\perp\perp}^2(1-\cos^2\theta^{\prime})\nonumber\\
&&\qquad\qquad\qquad\;\;\;\;+ \bar{F}_{\perp\perp}^{\cos\phip}\sin 2\theta^{\prime} \cos\phi^{\prime}+\bar{F}_{\perp\perp}^{\cos2\phip}\sin^2\theta^{\prime}\cos2\phi^{\prime}\Big]\nonumber\\
%
%sin phisa+phisb
&&\quad\;+S_{\perp A}^{\prime}S_{\perp B}^{\prime}\sin(\phi_{S_A}^{\prime}+\phi_{S_B}^{\prime})
\Big[F_{\perp \perp}^{\sin\phip}\sin 2\theta^{\prime}\sin\phi^{\prime}+F_{\perp \perp}^{\sin2\phip}\sin^2\theta^{\prime}\sin2\phi^{\prime}\Big]\nonumber\\
%sin phisa-phisb
&&\quad\;+S_{\perp A}^{\prime}S_{\perp B}^{\prime}\sin(\phi_{S_A}^{\prime}-\phi_{S_B}^{\prime})
\Big[\bar{F}_{\perp\perp}^{\sin\phip}\sin 2\theta^{\prime}\sin\phi^{\prime}+\bar{F}_{\perp\perp}^{\sin2\phip}\sin^2\theta^{\prime}\sin2\phi^{\prime}\Big]\nonumber\\
%
%LT
&&\quad\;+S_{\ell A}^{\prime}S_{\perp B}^{\prime}\cos\phi_{S_B}^{\prime}\Big[F_{\ell \perp}^1(1+\cos^2\theta^{\prime})+ F_{\ell \perp}^2(1-\cos^2\theta^{\prime})\nonumber\\
&&\qquad\qquad\qquad\;\;\;\;+ F_{\ell \perp}^{\cos\phip}\sin 2\theta^{\prime} \cos\phi^{\prime}+F_{\ell \perp}^{\cos2\phip}\sin^2\theta^{\prime}\cos2\phi^{\prime}\Big]\nonumber\\
&&\quad\;+S_{\ell A}^{\prime}S_{\perp B}^{\prime}\sin\phi_{S_B}^{\prime}
\Big[F_{\ell \perp}^{\sin\phip}\sin 2\theta^{\prime}\sin\phi^{\prime}+F_{\ell \perp}^{\sin2\phip}\sin^2\theta^{\prime}\sin2\phi^{\prime}\Big]\nonumber\\
%TL
&&\quad\;+S_{\perp A}^{\prime}S_{\ell A}^{\prime}\cos\phi_{S_A}^{\prime}\Big[F_{\perp\ell}^1(1+\cos^2\theta^{\prime})+ F_{\perp\ell}^2(1-\cos^2\theta^{\prime})\nonumber\\
&&\qquad\qquad\qquad\;\;\;\;+ F_{\perp\ell}^{\cos\phip}\sin 2\theta^{\prime} \cos\phi^{\prime}+F_{\perp\ell}^{\cos2\phip}\sin^2\theta^{\prime}\cos2\phi^{\prime}\Big]\nonumber\\
&&\quad\;+S_{\perp A}^{\prime}S_{\ell B}^{\prime}\sin\phi_{S_A}^{\prime}
\Big[F_{\perp\ell}^{\sin\phip}\sin 2\theta^{\prime}\sin\phi^{\prime}+F_{\perp\ell}^{\sin2\phip}\sin^2\theta^{\prime}\sin2\phi^{\prime}\Big]
\Bigg\}\,.
\label{eq:doublepxs}
\end{eqnarray}
We point out that, as for the case of single polarized Drell-Yan cross section, a transverse/longitudinal spin vector in the hadronic
c.m. frame acquires a longitudinal/transverse component in the dilepton rest frame, see Eqs.~(\ref{eq:stymu}-\ref{eq:stxmu}). However, as for the case of single polarized Drell-Yan cross section, in the CS frame at first order in a $q_T/M$ expansion Eq.~(\ref{eq:doublepxs}) holds with $\perp\equiv T$
and $\ell \equiv L$.
Eq.~(\ref{eq:doublepxs}) shows a considerably rich structure, which is related to all the possible correlations between the hadron and parton spins and the parton transverse momenta.
% It is interesting to notice that in the context of a collinear approximation, where transverse momenta are neglected, we would only have two contributions, corresponding to the terms proportional to the convolution of two transversity distribution functions and of two helicity distributions.

We can now conclude that the polarized hadronic tensor, Eq.~(\ref{eq:WSASB}),
can be decomposed in terms of 48 structure functions: 4 for $W^{\mu\nu}_{unp}$,
8 for $W^{\mu\nu}_{S_A}$, 8 for $W^{\mu\nu}_{S_B}$ and 28 for $W^{\mu\nu}_{S_A S_B}$ in agreement with Ref.~\cite{Arnold:2008kf}.

\section{Parton model}\label{P-M}

In the previous sections we analyzed the general structure of the Drell-Yan cross section in terms of hadronic structure functions. Here, we will explicitly calculate some terms of this cross section in the context of the parton model. Ref.~\cite{Arnold:2008kf} provides a full classification of the expressions of the (polarized and unpolarized) hadronic structure functions in the parton model at born level ($q\bar{q}$ electromagnetic annihilation), having taken into account the intrinsic motion of partons inside the nucleons.

Our aim is to show that the decomposition of the hadronic tensor $W^{\mu\nu}$ along the dilepton rest frame axes can also be easily applied to the partonic tensor, $w^{\mu\nu}$, in principle at any order in $q_T/M$ (although factorization is not  guaranteed at any order). At present, we will only consider two relevant examples at first order in $q_T/M$: the unpolarized parton distribution function contributions to the unpolarized cross section, and the Sivers contributions to the single polarized cross section, which originate from the {\it unpolarized}  partonic tensor.
The general expression for the polarized cross section, that requires the full structure of polarized partonic tensor, summed over all underlying polarized $q\bar{q} \to \ell ^+\ell ^-$  annihilations, will be presented in a forthcoming paper.

We work in a kinematical configuration in which hadrons $A$ and $B$ move along the $\Zcm$ axis, in the $A$-$B$ c.m. frame. Therefore, neglecting hadron masses, their four-momenta are:
\begin{equation}
p_A=\frac{\sqrt{s}}{2}\Big(1,0,0,1\Big)\,,
\,\,\,\, \,\,\,\,\,\, \,\,
p_B=\frac{\sqrt{s}}{2}\Big(1,0,0,-1\Big)\,,
\end{equation}
where $s\equiv(p_A+p_B)^2$. Taking into account the parton transverse motion, the partonic momenta
 can be written as:
\begin{eqnarray}
p_a&=&x_a\frac{\sqrt{s}}{2}\Big(
1+\frac{{k}^2_{\perp a}}{x_a^2 s},%E
\frac{2\,\bfk_{\perp a}}{x_a\sqrt{s}},%k
1-\frac{{k}^2_{\perp a}}{x_a^2 s}%z
\Big)\label{pa}\,,\\
p_b&=&x_b\frac{\sqrt{s}}{2}\Big(
1+\frac{{k}^2_{\perp b}}{x_b^2 s},%E
\frac{2\,\bfk_{\perp b}}{x_b\sqrt{s}},%k
-1+\frac{{k}^2_{\perp b}}{x_b^2 s}%z
\Big)\,,\label{pb}
\end{eqnarray}
where $x_a$ and $x_b$ denote the light-cone momentum fractions of parton $a$ and $b$ respectively.
The virtual photon $\gamma^*$ has a momentum $q$ which, by momentum conservation, equals the sum of the two partonic momenta $p_a$ and $p_b$:
\begin{eqnarray}
q&=&(p_a+p_b)=(q_0, \bfq_T,q_L)\,\qquad {\rm with}\qquad q^2=M^2\label{qdef}\,.
\end{eqnarray}
To ${\cal O}(q_T/M)$  we have:
\begin{eqnarray}
&&x_a\simeq\frac{q_{0}+q_L}{\sqrt{s}}\,,\,\,\,\, \,\,\,\,\,\, \,\,x_b\simeq\frac{q_{0}-q_L}{\sqrt{s}}\,.
\label{xaxbcoll}
\end{eqnarray}
which implies $x_a x_b\simeq M^2/s$.

According to the QCD parton model, the hadronic tensor amplitude $W^{\mu\nu}$ is the incoherent sum of the corresponding elementary partonic tensors for $q\bar q$ annihilations, $w^{\mu\nu}$.
Therefore, to leading order, the Drell-Yan cross section can be written as
\begin{equation}
\frac{d\sigma}{d^4 q d \Omega^{\prime}}=\frac{\alpha^2 }{6M^2 s} \sum_{q} e^2_q \int d^2 \bfk_{\perp a} \frac{L^{\mu\nu} \, w_{\mu\nu}}{M^4}\;f_{a/S_A}^q(x_a,\bfk_{\perp a})\; \bar f_{b/S_B}^{q} (x_b,\bfq_T-\bfk_{\perp a})
\label{parton-model-xs}\,,
\end{equation}
where $a,b=q,\bar q$ with $q=u, d, s,\bar u, \bar d,  \bar s$,
while $L^{\mu\nu}$ is the leptonic tensor given in Appendix~\ref{app:LepTens}. For the time being, we will only  consider the contribution of {\it unpolarized partons}, therefore the expression of the unpolarized partonic tensor in the dilepton rest frame is identical to that of the leptonic tensor, with $\phip$ and $\thetap$ replaced by $\phip_{a}$ and $\thetap_a$, the azimuthal and polar angles of parton $q$ in the dilepton rest frame
\begin{eqnarray}
w^{\mu\nu}&=&-\frac{M^2}{2}\Bigg\{\left(g^{\mu\nu}-\frac{q^{\mu}q^{\nu}}{M^2}+z^{\mu}z^{\nu}\right)(1+\cos^2\thetap _a)-2\,\sin^2\thetap _a \; z^{\mu}z^{\nu}\nonumber\\
&&\qquad+2\sin^2\thetap_a\cos2\phip_a\left[x^{\mu}x^{\nu}+\frac{1}{2}\left(g^{\mu\nu}-\frac{q^{\mu}q^{\nu}}{M^2}+z^{\mu}z^{\nu}\right)\right]
%\nonumber\\&&\qquad
+\,\sin^2\thetap_a\sin2\phip_a (x^{\mu}y^{\nu}+y^{\mu}x^{\nu})
\nonumber\\&&\qquad
+\sin2\thetap_a\cos\phip_a( x^{\mu}z^{\nu}+z^{\mu}x^{\nu})
%\nonumber\\&&\qquad
+\sin2\thetap_a\sin\phip_a ( y^{\mu}z^{\nu}+z^{\mu}y^{\nu})\Bigg\}\label{partonic_tensor}\,.
\end{eqnarray}
In Eq.~(\ref{parton-model-xs}), $f_{a/S_A}$ denotes the distribution function of an unpolarized parton of flavour $a$ inside a hadron $A$ with polarization $S_A$, and is given by the sum of the unpolarized distribution function, $f_{a/A}(x_a,k _{\perp a})$, and the Sivers distribution function, $\Delta^N f_{a/\uparrow}(x_a,\bfk_{\perp a})$, which gives the density number of an unpolarized parton $a$ inside a transversely polarized hadron $A$~\cite{Anselmino:2005sh, Anselmino:2011ch}:
% is related to the distribution of the same unpolarized parton inside the unpolarized hadron $A$ by
%
\begin{eqnarray}
f_{a/S_A}(x_a,\bfk_{\perp a})&=&f_{a/A}(x_a,k_{\perp a})+\frac{1}{2}\Delta^N f_{a/\uparrow}(x_a,k_{\perp a}) \sin(\phiSA-\phi_a)\nonumber\\
f_{b/S_B}(x_b,\bfk_{\perp b})&\Rightarrow&f_{b/B}(x_b,k_{\perp b})\,.
\label{distributions}
\end{eqnarray}
Notice that the Sivers function is also often denoted as $f_{1T}^{\perp a}(x_a, k_{\perp a})$;
this notation is related to ours by
\be
\Delta^N \! f_ {a/\uparrow}(x_a,k_{\perp a}) = - \frac{2\,k_{\perp a}}{m_p} \>
f_{1T}^{\perp a}(x_a, k_{\perp a}) \>. \label{rel}
\ee

The Sivers function is a particularly interesting TMD to unravel the 3-D structure of hadrons, as it embeds the coupling among the transverse motion of unpolarized partons and their parent hadron spin, which can only be related to their orbital angular momentum. Therefore, a clean extraction of the Sivers function will also shed light on the angular momentum of quarks and gluons inside hadrons. Moreover,
it is now widely  accepted that this function should change sign~\cite{Collins:2004nx} when going from SIDIS to Drell Yan processes, a belief that still needs to be supported by experimental evidence. It is therefore of great interest to be able to perform a clear and exhaustive analysis of its contribution to Drell Yan processes: although this is extremely involved in the general case, as shown in Eqs.~(\ref{eq:singlepxs2}), it can become particularly simple and manifest when the process is examined in the kinematical range in which the photon transverse momentum, $q_T$, is much smaller than the dilepton invariant mass, $M$, and of the same order of magnitude of the quark and anti-quark intrinsic momenta, as we will illustrate in what follows.

Contracting the unpolarized leptonic and partonic tensors we obtain:
\begin{eqnarray}
 L^{\mu\nu} w_{\mu\nu}&=&\frac{M^4}{2}\Big\{(1+\cos^2\theta^{\prime})(1+\cos^2\theta^{\prime}_a)+2\,\sin^2\theta^{\prime}\sin^2\theta^{\prime}_a\nonumber\\
&&\quad \quad\;\;+ 4\,\sin^2\thetap\cos2\phip\sin^2\thetap_a\cos2\phip_a\nonumber\\
&&\quad \quad\;\;+\sin^2\thetap\sin2\phip\sin^2\thetap_a\sin2\phip_a\nonumber\\
&&\quad \quad\;\;+ \sin 2\thetap\cos\phip\sin2\thetap_a\cos\phip_a\nonumber\\
&&\quad \quad\;\;+\sin2\thetap\sin\phip\sin2\thetap_a\sin\phip_a\Big\}\label{eq:lwcontracted}\,.
\end{eqnarray}
Inverting Eq.~(\ref{eq:lambdacs}) we can connect the components of partonic momenta in the dilepton rest frame with those in the hadronic c.m. frame; to first in order $q_T/M$ we find:
\begin{eqnarray}
\cos^2\thetap_a&\simeq& 1 \label{cos} \\
\sin^2\thetap_a&\simeq& 0 \label{sin} \\
\sin 2\thetap_a \cos\phi_a&\simeq&-\frac{2}{M}(q_T-2 \kax\cos{\phig}-2 \kay\sin{\phig})\\
 \sin 2\thetap_a \sin\phi_a&\simeq&-\frac{4}{M}(\kax\sin{\phig}- \kay\cos{\phig})\,,
\end{eqnarray}
where $\kax$ and $\kay$ are the transverse components of $p_a$ in the hadronic c.m. frame. Eqs.~(\ref{cos}) and (\ref{sin}) tell us that only three leading terms survive in Eq.~(\ref{eq:lwcontracted}). Thus:
\begin{eqnarray}
 L^{\mu\nu} \omega_{\mu\nu}&\simeq&M^4\Bigg\{(1+\cos^2\theta^{\prime})\nonumber\\
&&\quad \quad\;\;-\,\sin 2\thetap\cos\phip\left(\frac{q_T}{M}- \frac{2k_{\perp a 1}}{M}\cos{\phig}- \frac{2k_{\perp a 2}}{M}\sin{\phig}\right)\nonumber\\
&&\quad \quad\;\;-\sin2\thetap\sin\phip\left( \frac{2k_{\perp a 1}}{M}\sin{\phig}- \frac{2k_{\perp a 2}}{M}\cos{\phig}\right)\Bigg	\}\,.\label{eq:lwapx}
\end{eqnarray}
At this stage, the final expression for the cross section is obtained by inserting Eqs.~(\ref{distributions}) and (\ref{eq:lwapx}) into Eq.~(\ref{parton-model-xs}). The integration on $\bfk_{\perp a}$ can be further simplified by reconsidering the integrals after a rotation of the system to the ``virtual photon'' production plane, as described in detail in Appendix~\ref{int-rot}. Finally we find:
\begin{eqnarray}
\frac{d\sigma}{d^4 q d \Omega^{\prime}}&=&
\frac{\alpha^2}{6 M^2 s} \sum_{q} e^2_q
\left\{(1+\cos^2\thetap)\int\! d^2 \bfk_{\perp a}
\,f^q_{a/A}(x_a,k_{\perp a})\bar f^q_{b/B}(x_b,k_{\perp b})\right. \nonumber \\
&&\hspace*{-1.2cm}\left.+
\sin 2\thetap \cos \phip\int\! d^2 \bfk_{\perp a}
\,\left[ 2\frac{k_{\perp a}}{M} (\hat{\bfk}_{\perp a}\cdot\hat{\bfq}_T)-\frac{q_T}{M} \right]
f^q_{a/A}(x_a,k_{\perp a})\bar f^q_{b/B}(x_b,k_{\perp b})\right. \nonumber \\
&&\hspace*{-1.2cm}\left.+ (1+\cos^2\thetap)\sin(\phi_{S_A}-\phig)\int \!d^2 \bfk_{\perp a}
(\hat{\bfk}_{\perp a}\cdot\hat{\bfq}_T)\Delta^N f^q_{a/\uparrow}(x_a,k_{\perp a})\bar f^q_{b/B}(x_b,k_{\perp b})\right. \nonumber \\
&&\hspace*{-1.2cm}\left.+\sin 2\thetap \cos \phip \sin(\phi_{S_A}-\phig)
\!\int\! d^2\bfk_{\perp a}
\,(\hat{\bfk}_{\perp a}\cdot\hat{\bfq}_T)\left[ 2\frac{k_{\perp a}}{M} (\hat{\bfk}_{\perp a}\cdot\hat{\bfq}_T)-\frac{q_T}{M}  \right]\Delta^N f^q_{a/\uparrow}(x_a,k_{\perp a})\bar f^q_{b/B}(x_b,k_{\perp b})\right. \nonumber \\
&&\hspace*{-1.2cm}\left.+ \sin 2\thetap \sin \phip \cos(\phi_{S_A}-\phig)
\!\int\! d^2\bfk_{\perp a} \;2\frac{k_{\perp a}}{M}\left[ (\hat{\bfk}_{\perp a}\cdot\hat{\bfq}_T)^2-1  \right]\,\Delta^N f^q_{a/\uparrow}(x_a,k_{\perp a})\bar f^q_{b/B}(x_b,k_{\perp b})
\right\}\,,
\label{final-xs}
\end{eqnarray}
where we recall that the lepton angles $\thetap$ and $\phip$ are expressed in the CS frame, while $\phi_{S_A}$ and $\phig$ are referred to the hadronic c.m. frame.
This relation deserves a careful discussion.
\begin{itemize}
\item
In the first line we find the leading term of the unpolarized cross section, proportional to the convolution of the two unpolarized parton distribution functions, $f_{a/A}(x_a,k_{\perp a}) \otimes f_{b/B}(x_b,k_{\perp b})$, and identified by the $(1+\cos^2\thetap)$ angular distribution.
\item
In the second line, we find the first order correction to the previous term in our $q_T/M$ expansion (notice that $k_\perp/M$ and $q_T/M$ are of the same order). This term can be considered
as the analogous of the so called Cahn effect in SIDIS~\cite{Cahn:1978se,Cahn:1989yf}. In Eq.~(\ref{C9}) of Appendix~\ref{int-rot} we show that it is directly proportional to the difference between the mean average transverse momenta of parton $a$ and of parton $b$ inside the two colliding hadrons.
In the Gaussian model hypothesis, see Eq.~(\ref{unp-gaussian}),
this term translates directly in the difference between the Gaussian widths.
Thus such term is negligible when the two colliding hadrons are charge conjugated, as it is the case for $p\bar p$ induced DY processes, while it could be observed in $\pi p$ DY processes if the pion and proton transverse momentum distribution are different, or in $pp$ scattering if the quark transverse momentum distribution (valence contribution) is different from that of the anti-quark (sea contribution).
Although the existence of this term in the unpolarized cross section was pointed out many years ago in Refs.~\cite{Lam:1978pu} and~\cite{Collins:1977iv}, it is only now, in light of the new theoretical developments and of the planned experimental programs, that its phenomenological implications related to its detailed form are becoming particularly relevant.
Finally, we point out that here we do not find the term proportional to $\sin ^2\thetap \cos 2\phip$, i.e. the  so-called Boer-Mulders term, which is usually included in the unpolarized DY cross section. This is due to the fact that we are not considering the contributions originating from any polarized elementary $q\bar q$ annihilation. Its absence, however, does not invalidate our phenomenological considerations, as this term gives rise to a different azimuthal dependence and can, therefore, be experimentally isolated.
\item
The last three lines correspond to the contributions of polarized hadrons to the cross section,
and involve the convolution of the Sivers distribution function $\Delta^N f_{a/\uparrow}(x_a,k_{\perp a})$ with the unpolarized distribution function $f_{b/B}(x_b,k_{\perp b})$: line three corresponds to the dominant Sivers effect which, in DY processes, is of great importance from the theoretical point of view.
As mentioned above, the Sivers function reverses its sign when is observed in DY and SIDIS processes~\cite{Collins:2004nx}: 
presently,  no  polarized DY experiments are  able to confirm this prediction but many dedicated future experiments are planned in several facilities: RHIC at BNL~\cite{Bunce:2000uv}, PANDA~\cite{Brinkmann:2007zz} and PAX~\cite{Barone:2005pu} at FAIR, COMPASS at CERN~\cite{Gautheron:1265628}, IHEP~\cite{Vasiliev:2007nz} in Protvino, and JINR~\cite{NICA} in Dubna.
\item
The last two lines of Eq.~(\ref{final-xs}), which can be rewritten in an alternative, more suitable form as
\begin{eqnarray}
&...&\sin 2\thetap\sin(\phiSA-\phig+\phip)\int d^2\bfk_{\perp a}\frac{1}{2M}
\Big[4k_{\perp a}(\hat{\bfk}_{\perp a}\cdot \hat{\bfq}_T)^2+ \hat{\bfk}_{\perp a}\cdot(\bfq_T-2 \bfk_{\perp a})\Big]\,\Delta^N f^q_{a/\uparrow} \; \bar f^q_{b/B}\nonumber\\
&+&\sin 2\thetap\sin(\phiSA-\phig-\phip)
\int d^2\bfk_{\perp a}\frac{2k_{\perp a}-q_T ( \hat{\bfk}_{\perp a}\cdot\hat{\bfq}_T )}{2M}\,
\Delta^N f^q_{a/\uparrow} \; \bar f^q_{b/B}\,,
\end{eqnarray}
represent  the equivalent of the ``Cahn effect'' for the single polarized DY, and show a structure very similar to that of the unpolarized part.
These contributions are suppressed  by one power of $q_T/M$ and, as in the unpolarized Cahn effect,
we can recast the integration on $\bfk_{\perp a}$ similarly to what is done in Eq.~(\ref{C9}): by doing this one can observe that they can give access to the difference between the average transverse momenta of the unpolarized and the Sivers distribution functions.
\end{itemize}
These results become particularly transparent and acquire a phenomenological value when the integrals over the intrinsic transverse momenta are explicitly performed by using a simple Gaussian model for the TMDs. Similarly to what was done in the last Section of Ref.~\cite{Anselmino:2011ch}, we assume the $k_\perp$ dependence of the TMDs can be factored and approximated with a Gaussian distribution of the form:
\be
f_{a/A} (x_a,k_{\perp a})= f_{a/A} (x_a)\,\frac{e^{-k_{\perp a}^2/\langle k_{\perp a}^2\rangle}}{\pi\langle k_{\perp a}^2\rangle }\,,
\label{unp-dist}\\
\ee
where $f_{a/A} (x_a)$, can be taken
from the available fits of the world data. In general, we allow for
different widths of the Gaussians for the different parton flavours, but take
them to be constant. For the Sivers
function, we assume a similar parametrization, with an extra multiplicative
factor $k_{\perp a}$ to give it the appropriate behavior in the small $k_{\perp a}$ region~\cite{Anselmino:2008sga}:
\be
\Delta f_{a/\uparrow} (x_a, k_{\perp a}) = \Delta f_{a/\uparrow}(x) \; \sqrt{2e}\,\frac{k_{\perp a}}{M\S} \;
\frac{e^{-k_{\perp a}^2/\langle k_{\perp a}^2\rangle \S}}{\pi\langle k_{\perp a}^2\rangle}\,,
\label{Siv-dist}
\ee
where the $x$-dependent function $\Delta f_{a/\uparrow}(x_a)$
is not known, and should be determined
phenomenologically by fitting the available data on azimuthal asymmetries and
moments; the $k_\perp$ dependent Gaussian has been assigned a width $\langle k_{\perp a}^2\rangle \S$ and a suitable normalization coefficient $\sqrt{2e}$
to make sure it fulfills the appropriate positivity bounds~\cite{Bacchetta:1999kz}.

By inserting Eqs.~(\ref{unp-dist}) and (\ref{Siv-dist}) into Eq.~(\ref{final-xs}) we get
\begin{eqnarray}
\frac{d\sigma^{unp}}{d^4 q d\Omega^{\prime}}=
\frac{\alpha^2}{6 M^2 s}\sum_{q} e_q^2 \;f_{a/A}^q(x_a)\;\bar{f}_{b/B}^q(x_b)
\frac{e^{-q_T^2/\langle q_T ^2\rangle }}{\pi\langle q_T ^2\rangle}\;
\Big\{(1+\cos^2\thetap)+
\frac{q_T}{M}\;
\frac{\langle k_{\perp a}^2\rangle-\langle k_{\perp b}^2\rangle}{\langle q_T ^2\rangle}\sin2\thetap\cos \phip\;
\Big\}\;,\nonumber \\ \label{unp-gaussian}
\end{eqnarray}
and
\begin{eqnarray}
\frac{d\sigma^{S_A}}{d^4 q d\Omega^{\prime}}&=&
\frac{\sqrt{2e} \,\alpha^2 }{12 M^2 s}\;
\frac{q_T}{M_{S}}
\sum_{q} e_q^2\;\Delta f_{a/\uparrow}^q(x_a)\;\bar{f}_{b/B}^q(x_b)\;
\frac{e^{-q_T^2/\langle q_T ^2\rangle\S}}{\pi\langle q_T ^2\rangle\S ^2}\;
\frac{\kpss\S^2}{\langle k_{\perp a}^2\rangle}\;\times \nonumber\\
&&\Big\{(1+\cos\thetap)\sin(\phi_{S_A}-\phi_{\gamma})\nonumber\\
&&+
\frac{q_T}{M}\Big[\Big(
\frac{\kpss\S-\langle k_{\perp b}^2\rangle}{\langle q_T ^2\rangle \S}
+\frac{\langle k_{\perp b}^2\rangle}{ q^2_T}\Big)\; \sin2\thetap\cos\phip \sin(\phi_{S_A}-\phi_{\gamma})
\Big.\nonumber\\&&\Big.
-\frac{\langle k_{\perp b}^2\rangle}{q_T^2}\;\sin2\thetap\sin\phip \cos(\phi_{S_A}-\phi_{\gamma})\Big]\Big\}\;,
\label{Siv-gauss}
\end{eqnarray}
where we have defined \(\langle k_{\perp a}^2 \rangle + \langle k_{\perp b}^2 \rangle \equiv \langle q_T ^2\rangle \), and \(\kpss\S^2 + \langle k_{\perp b}^2 \rangle \equiv \langle q_T ^2\rangle \S\).
Notice that Eq.~(\ref{Siv-gauss}) can be rearranged as Eq.~(\ref{eq:singlepxs2}):
\begin{eqnarray}
\frac{d\sigma^{S_A}}{d^4 q d\Omega^{\prime}}&=&
\frac{\sqrt{2e} \,\alpha^2 }{12 M^2 s}\;
\frac{q_T}{M_{S}}
\sum_{q} e_q^2\;
\Delta f_{a/\uparrow}^q(x_a)\;\bar{f}_{b/B}^q(x_b)\;
\frac{e^{-q_T^2/\langle q_T ^2\rangle \S}}
{\pi\left(\langle q_T ^2\rangle \S\right)^2}\;
\frac{\kpss\S ^2}{\langle k_{\perp a}^2\rangle}\;\;\times \nonumber\\
&&\Big\{(1+\cos^2\thetap)\sin(\phi_{S_A}-\phi_{\gamma})\nonumber\\
&&+\sin2\thetap\sin(\phiSA-\phig+\phip)\Big[\frac{q_T}{2M}
\frac{\kpss\S-\langle k_{\perp b}^2\rangle}{\langle q_T ^2\rangle \S}
\Big]\nonumber\\
&&+\frac{1}{2}\sin 2\thetap\sin(\phiSA-\phig-\phip)\;\Big[\frac{\kpss\S-\langle k_{\perp b}^2\rangle}{\langle q_T ^2\rangle \S}
+\frac{2 k_{\perp b}^2}{M q_T}\Big]\Big\}\,.
\label{Siv-gauss-2}
\end{eqnarray}
Eqs.~(\ref{Siv-gauss}) and (\ref{Siv-gauss-2}) have a great phenomenological value as they show that if a $\sin(\phiSA-\phig+\phip)$ azimuthal modulation is experimentally detected, then it gives direct access to the asymmetry $\kpss\S-\langle k_{\perp b}^2\rangle$, whereas the $\sin \phip\cos(\phiSA-\phig)$ azimuthal modulation would give access to $\langle k_{\perp b}^2\rangle$ alone, and would survive even in the case in which $\kpss\S=\langle k_{\perp b}^2\rangle$.

Notice that, from SIDIS experiments we know that $\langle k_{\perp }^2\rangle\S \neq\langle k_{\perp }^2\rangle$:
in particular, in the latest data fits~\cite{Anselmino:2008sga,Anselmino:2010bs} the ratio
$\langle k_{\perp }^2\rangle\S/\langle k_{\perp }^2\rangle$ is rather well constrained and turns out to be between one half and two thirds.
Consequently, the scenario of equal $k_\perp$ distributions for the unpolarized and Sivers distribution functions seems to be excluded, making even more interesting any further investigations in DY processes, where we could have access to the values of  $\langle k_{\perp }^2\rangle\S$ and $\langle k_{\perp }^2\rangle$ separately.

\section{Conclusions}\label{CONCL}

The theoretical and phenomenological relevance that Drell-Yan processes
have recently assumed in the context of hadron physics
is out of any doubt.
New dedicated experiments are either under construction or being planned, like RHIC at BNL~\cite{Bunce:2000uv}, PANDA~\cite{Brinkmann:2007zz} and PAX~\cite{Barone:2005pu} at FAIR, COMPASS at CERN~\cite{Gautheron:1265628}, IHEP~\cite{Vasiliev:2007nz} in Protvino, and JINR~\cite{NICA} in Dubna.
They will help to shed light  on many issues,
in particular on our understanding of the 3-D dynamical structure of nuclei and,
more in general, of hadrons in terms of their elementary constituents,
quarks and gluons.
This motivates our efforts toward the full understanding of the DY general cross sections, understanding that is mainly achieved by the decomposition of the hadronic tensor in terms
of structure functions.

We have performed this decomposition by using the helicity structure functions, i.e.~projecting the hadronic tensor along the dilepton helicity axes.
Some work was already done
in this direction~\cite{Tangerman:1994eh,Pire:1983tv,Arnold:2008kf}:
in particular,
in Ref.~\cite{Arnold:2008kf}
the hadronic tensor was decomposed with a tensorial basis constructed with 
the dilepton momentum, the hadronic momenta and the spin vectors,
which resulted in a decomposition in terms of invariant structure functions.
Although this basis leads to the desired results,
it is not ideal to perform such a decomposition, as it generates 
a huge number of non-independent structure functions
that must subsequently be eliminated by exploiting all the symmetry
properties the hadronic tensor should exhibit. 

In our work we decompose the hadronic tensor
using the dilepton helicity axis as tensor basis, a choice which ensures that all the symmetry properties are fulfilled from the beginning, avoiding the initial proliferation of non-independent structure functions. 
Moreover, the helicity structure functions can be intimately related
to the elementary scattering process, being connected to the polarization status of the virtual photon.
%, and allow an immediate partonic interpretation.
Therefore, we believe we have found a suitable
and elegant working setup which enables us to disentangle
the $48$ structure functions of the polarized DY cross section
in the most general way,
in a configuration valid in any dilepton rest frame,
with a minimal amount of effort.

Finally, within the QCD parton model approximation,
we have performed a phenomenological analysis of the unpolarized term
of the DY cross section and of the Sivers effect,
both originating from the elementary scattering
of an {\it unpolarized} $q\bar q$ pair,
in the kinematical range in which $k_\perp \simeq q_T \ll M$.
Interesting results,
which could help gathering valuable information on the $k_\perp$
distribution of the unpolarized TMD distribution function,
$f_{q/p}(x_q,k_\perp)$, and of the Sivers TMD distribution function,
$\Delta ^N f_{q/p^\uparrow}(x_q,\bfk _\perp)$, are presented.

The $q_T/M$ suppressed terms in Eq.~(\ref{final-xs})  deserve a similar  phenomenological analysis, 
that we are planning to perform in a forthcoming paper.
%Such kind of terms are suppressed as $q_T/M$ therefore 
One can also wonder
if they could be studied above the so-called safe region, where the suppression should be less pronounced: in principle, if in a single polarized DY process the $J/\psi$ is found to be transversely polarized (in the CS frame)
one can apply Eq.~(\ref{final-xs}),
with obvious modifications to accommodate for the $J/\psi$ coupling,
in analogy to what has been done in Ref.~\cite{Anselmino:2004ki}.
A further extension of this work, including the evaluation of the terms generated by 
the transverse polarization of partons,
therefore related to the Boer-Mulders and transversity functions, will soon be presented.

\begin{acknowledgments}
The authors would like to thank Mauro Anselmino for many useful discussions and for his support throughout this work.
We acknowledge support of the European Community - Research Infrastructure
Activity under the FP7 ``Structuring the European Research Area''
program (HadronPhysics2, Grant agreement 227431), and partial support by MIUR under Cofinanziamento PRIN 2008.
\end{acknowledgments}

\appendix
\section{Leptonic Tensor}\label{app:LepTens}
For reference we report the structure of the unpolarized leptonic tensor decomposed by means of the helicity axes, see Ref.~\cite{Tangerman:1994eh}. It can be easily obtained by inserting the expressions of $l^\mu_3$ and $l^\nu_4$ in terms of the helicity axes into Eq.~(\ref{leptonic_tensor_master})
\begin{eqnarray}
l^\mu_3 &=& \frac{1}{2}\left[ q^\mu +M(\sin\thetap\cos\phip x^\mu + \sin\thetap\sin\phip y^\mu + \cos\thetap z^\mu) \right] \nonumber \\
l^\mu_4 &=& \frac{1}{2}\left[ q^\mu -M(\sin\thetap\cos\phip x^\mu + \sin\thetap\sin\phip y^\mu + \cos\thetap z^\mu) \right]\\
\end{eqnarray}
and keeping in mind that
\begin{equation}
g^{\mu\nu}=\frac{q^\mu q^\nu}{M^2}-x^\mu x^\mu-y^\mu y^\mu-z^\mu z^\mu
\end{equation}
as $x^\mu$, $y^\mu$ and $z^\mu$ are all perpendicular to $q^\mu$.
Thus one obtains:
\begin{eqnarray}
L^{\mu\nu}&=&-\frac{M^2}{2}\Bigg\{\left(g^{\mu\nu}-\frac{q^{\mu}q^{\nu}}{M^2}+z^{\mu}z^{\nu}\right)(1+\cos^2\thetap)-2\,\sin^2\thetap z^{\mu}z^{\nu}\nonumber\\
&&\qquad+2\sin^2\thetap\cos2\phip\left[x^{\mu}x^{\nu}+\frac{1}{2}\left(g^{\mu\nu}-\frac{q^{\mu}q^{\nu}}{M^2}+z^{\mu}z^{\nu}\right)\right]\nonumber\\
&&\qquad+\,\sin^2\thetap\sin2\phip (x^{\mu}y^{\nu}+y^{\mu}x^{\nu})+\sin2\thetap\cos\phip( x^{\mu}z^{\nu}+z^{\mu}x^{\nu})\nonumber\\
&&\qquad+\sin2\thetap\sin\phip ( y^{\mu}z^{\nu}+z^{\mu}y^{\nu})\Bigg\}\,.\label{leptonic_tensor}
\end{eqnarray}

\section{Collins-Soper helicity frame}\label{app-cs}

The leptonic angular distribution is studied in the dilepton rest frame where leptons are back to back. The dilepton rest frame can be chosen in many different ways.
A very common choice is the so called Collins-Soper (CS) frame.~\cite{Collins:1977iv}.
In the hadronic c.m. frame the CS axes read:
\begin{eqnarray}
 z^{CS}&=&\frac{1}{\sqrt{M^2+q_T^2}}\Big(q_L,0,0,q_0\Big)\\
x^{CS}&=&\frac{1}{\sqrt{M^2+q_T^2}}\Big(\frac{q_0 q_T}{M},\frac{M^2+q_T^2}{M}\cos\phi_{\gamma},\frac{M^2+q_T^2}{M}\sin\phi_{\gamma},\frac{q_L q_T}{M} \Big)\\
y^{CS}&=&\Big(0,-\sin\phi_{\gamma},\cos\phi_{\gamma},0\Big)
\end{eqnarray}
where $q^{\mu}=(q_0, q_T\cos\phi_{\gamma},q_T\sin\phi_{\gamma},q_L)$ is the dilepton (virtual photon) momentum in the hadronic $c.m.$ frame. Notice that the $y^{CS}$ axis has no time-like nor longitudinal components.
\\
The matrix:
\begin{equation}
\Lambda^{\mu\nu}_{CS}=
\left(\begin{array}{cccc}
\frac{q_0}{M} &\frac{q_0 q_T}{M\sqrt{M^2+q_T^2}}&0&\frac{q_L}{\sqrt{M^2+q_T^2}}\\
\frac{q_T \cos\phi_{\gamma}}{M}&\frac{\sqrt{M^2+q_T^2}}{M}\cos\phi_{\gamma}&-\sin\phi_{\gamma}&0\\
\frac{q_T\sin\phi_{\gamma}}{M}&\frac{\sqrt{M^2+q_T^2}}{M}\sin\phi_{\gamma}&\cos\phi_{\gamma}&0\\
\frac{q_L}{M}&\frac{q_L q_T}{M\sqrt{M^2+q_T^2}}&0&\frac{q_0}{\sqrt{M^2+q_T^2}}
\end{array}\right)\label{eq:lambdacs}
\end{equation}
is the Lorentz transformation from the Collins-Soper frame to the hadronic c.m. frame. Notice that the columns of this matrix are the four column vectors $t^{\mu}\equiv q^{\mu}/M$, $x^{\mu}$, $y^{\mu}$, $z^{\mu}$ in the Collins-Soper frame.

\section{Integration by rotation in the virtual photon production plane}\label{int-rot}

Let us define as the ``virtual photon'' production plane the plane containing the virtual photon momentum $\bfq$ and the $\Zcm$ axis in the hadron $c.m.$ frame. We can then define a new frame where $\Zg\equiv \Zcm$ and the $\Xg$ axes lay in the virtual photon production plane and $\Yg$ is perpendicular to both $\Xg$ and $\Zg$.
This new frame is rotated by an angle $\phig$ with respect to the c.m. frame $(\Xcm, \Ycm, \Zcm)$ :
\begin{eqnarray}
\Xcm&=&\Xg \cos\phig - \Yg \sin\phig\\
 \Ycm&=&\Xg \sin\phig + \Yg \cos\phig\\
\Zcm&\equiv&\Zg
\end{eqnarray}
As a direct consequence, the transverse photon momentum will be along $\Xg\equiv \hat{\bfq}_T$.
Notice also that, for partonic momentum conservation, any function of the parton intrinsic transverse momentum moduli, $g(k_{\perp a}^2,k_{\perp b}^2)$, obeys the relation $g(k_{\perp a}^2,k_{\perp b}^2)=g[k_{\perp a}, (\bfq_T-\bfk_{\perp a})^2]=\gkk$.
When we insert Eq.~(\ref{eq:lwapx}) into Eq.~(\ref{parton-model-xs}) the first two terms, proportional to the convolution of the unpolarized distribution functions $f_{a/A}\,f_{b/B} $, will show integrals of the form:
\begin{eqnarray}
&&\int  d^2 \bfk_{\perp a} \, \gkk (\kax \cos\phig+\kay\sin\phig)\nonumber\\
&&\qquad\qquad=\int  d^2 \bfk_{\perp a}  \,\gkk [(\bfk_{\perp a}\cdot \Xcm) \cos\phig+(\bfk_{\perp a}\cdot \Ycm)\sin\phig]\nonumber\\
&&\qquad\qquad=\int  d^2 \bfk_{\perp a} \, \gkk(\bfk_{\perp a}\cdot \Xg)\,,
\end{eqnarray}
\begin{eqnarray}
\int  d^2 \bfk_{\perp a}  \,\gkk (\kax \sin\phig-\kay\cos\phig)&=&
-\int  d^2 \bfk_{\perp a}  \,\gkk(\bfk_{\perp a}\cdot \Yg)=0\,,\nonumber \\
\end{eqnarray}
having exploited the fact that $\int  d^2 \bfk_{\perp a}  \,\gkk(\bfk_{\perp a}\cdot \Yg)=0$, because $(\bfk_{\perp a}\cdot \Yg)$ is an odd function over the symmetric range of $\bfk_{\perp a}$ integration.

The Sivers terms are more involved, as the Sivers function itself depends on $\bfk_{\perp a}$ through its proper phase, $\sin(\phi_{S_A} - \phi_{\perp a}) \propto (\sin\phi_{S_A} \,\kax - \cos\phi_{S_A} \,\kay) $. Therefore, we
will have to deal with integrals of the form
\begin{eqnarray}
&&\int  d^2 \bfk_{\perp a}  \,\gkk (\kax \sin\phiSA -\kay\cos\phiSA)\nonumber \\
&&\qquad\qquad\qquad=\int  d^2 \bfk_{\perp a} \, \gkk[(\bfk_{\perp a}\cdot \Xg)\sin(\phiSA-\phig)
-(\bfk_{\perp a}\cdot \Yg)\cos(\phiSA-\phig)]\nonumber\\
&&\qquad\qquad\qquad=\int  d^2 \bfk_{\perp a} \, \gkk(\bfk_{\perp a}\cdot \Xg)\sin(\phiSA-\phig)\,,
\end{eqnarray}
\begin{eqnarray}
\int  d^2 \bfk_{\perp a}  \,\gkk (\kax \cos\phig+\kay\sin\phig)(\kax \sin\phiSA -\kay\cos\phiSA)\nonumber\\
=\int  d^2 \bfk_{\perp a}  \,\gkk (\bfk_{\perp a}\cdot \Xg)^2\sin(\phiSA-\phig)\,,
\end{eqnarray}
\begin{eqnarray}
\int  d^2 \bfk_{\perp a} \, \gkk  (\kax \sin\phig-\kay\cos\phig)(\kax \sin\phiSA -\kay\cos\phiSA)\nonumber\\
=\int  d^2 \bfk_{\perp a}  \,\gkk [k_{\perp a}^2-(\bfk_{\perp a}\cdot \Xg)^2]\cos(\phiSA-\phig)\,.
\end{eqnarray}

Finally, it is very interesting to notice that
\begin{eqnarray}
&&\int  d^2 \bfk_{\perp a}\, \, \gkk[q_T-2(\bfk_{\perp a}\cdot \Xg)]\nonumber\\
&&\qquad=\int d^2 \bfk_{\perp a} d^2\bfk_{\perp b} \delta^2(\bfk_{\perp a}+\bfk_{\perp q}-\bfq_T  )
g(k_{\perp a}^2,k_{\perp b}^2)\frac{1}{q_T}[(\bfk_{\perp b}-\bfk_{\perp a})\cdot (\bfk_{\perp b}+\bfk_{\perp a}) ]\nonumber\\
&&\qquad=\int d^2 \bfk_{\perp a} d^2\bfk_{\perp b} \delta^2(\bfk_{\perp a}+\bfk_{\perp q}-\bfq_T  )
g(k_{\perp a}^2,k_{\perp b}^2)\frac{k_{\perp b}^2-k_{\perp a}^2}{q_T}\,,
\label{C9}
\end{eqnarray}
which lead to the conclusions outlined in Sect.~\ref{P-M}

\bibliographystyle{h-physrev5-ste}
\bibliography{biblio2}

\end{document}